\documentclass[twoside,english,reprint, aps, prx, superscriptaddress, amsfonts, amssymb, amsmath, a4paper]{revtex4-1}
\usepackage{lmodern}

\usepackage[T1]{fontenc}
\usepackage[latin9]{inputenc}
\usepackage{color}
\usepackage{babel}
\usepackage{amsmath}
\usepackage{amssymb}
\usepackage{graphicx}
\usepackage[unicode=true,
 bookmarks=false,
 breaklinks=true,pdfborder={0 0 0},pdfborderstyle={},backref=false,colorlinks=true]
 {hyperref}
\hypersetup{pdftitle={Quantum tomography of entangled spin-multi-photon states},
 pdfauthor={Dan Cogan}}



\begin{document}
\title{Quantum tomography of entangled spin-multi-photon states.}
\author{Dan Cogan}
\affiliation{The Physics Department and the Solid State Institute, Technion\textendash Israel
Institute of Technology, 3200003 Haifa, Israel}
\author{Giora Peniakov}
\affiliation{The Physics Department and the Solid State Institute, Technion\textendash Israel
Institute of Technology, 3200003 Haifa, Israel}
\author{Oded Kenneth}
\affiliation{The Physics Department and the Solid State Institute, Technion\textendash Israel
Institute of Technology, 3200003 Haifa, Israel}
\author{Yaroslav Don}
\affiliation{The Physics Department and the Solid State Institute, Technion\textendash Israel
Institute of Technology, 3200003 Haifa, Israel}
\author{David Gershoni}
\email{dg@physics.technion.ac.il}

\affiliation{The Physics Department and the Solid State Institute, Technion\textendash Israel
Institute of Technology, 3200003 Haifa, Israel}
\begin{abstract}
We present a novel method for quantum tomography of multi-qubit states.
We apply the method to spin-multi-photon states, which we produce
by periodic excitation of a semiconductor quantum-dot- confined spin
every 1/4 of its coherent precession period. These timed excitations
lead to the deterministic generation of strings of entangled photons
in a cluster state. We show that our method can be used for characterizing
the periodic process map, which produces the photonic cluster. From
the measured process map, we quantify the robustness of the entanglement
in the cluster. The 3-fold enhanced generation rate over previous
demonstrations reduces the spin decoherence between the pulses and
thereby increases the entanglement.
\end{abstract}
\maketitle
\global\long\def\ket#1{\left|#1\right\rangle }%
\global\long\def\im{\operatorname{Im}}%
\global\long\def\bra#1{\left\langle #1\right|}%

\section{Introduction}

Measurement-based quantum protocols are very promising for quantum
computation in general \citep{Raussendorf2001,Raussendorf2003,Briegel2009}
and for quantum communication in particular \citep{Briegel1998,Zwerger2012,Zwerger2013,Azuma2015}.
The use of multi-partite entangled state named graph-states \citep{Briegel2001,Hein2004},
enables quantum computation by single qubit measurements and rapid
classical feedforward, depending on the measurement outcome \citep{Raussendorf2003}.
For quantum communication, graph-states of photons are particularly
attractive \citep{Walther_2005,Prevedel_2007,Lu_2007,Tokunaga_2008},
since they provide redundancy against photon loss, and compensation
for the finite efficiency of quantum gates. Moreover, since the quantum
information is contained in the graph state they eliminate the need
to communicate within the coherence time of the local nodes \citep{Azuma2015}.
Graph-states are therefore considered for efficient distribution of
entanglement between remote nodes \citep{Kimble2008} as well as for
quantum repeaters \citep{Zwerger2012,Zwerger2013}. Developing devices
capable of deterministically producing high-quality photonic graph
states at a fast rate is, therefore, a scientific and technological
challenge of utmost importance \citep{Munro2012}.

The technological quest for generating photonic graph states which
are required for building scalable quantum network architectures,
led to new schemes. Of particular importance and relevance to this
work is the Lindner and Rudolph proposal \citep{Lindner2009} for
generating one-dimensional cluster state of entangled photons using
semiconductor quantum dots (QDs). The scheme uses a single confined
electronic spin in a coherent superposition of its two eigenstates.
The spin precesses in a magnetic field while driven by a temporal
sequence of resonant laser pulses. Upon excitation of the QD spin,
a single photon is deterministically emitted and the photon polarization
is entangled with the polarization of the QD spin. This timed excitation
repeats itself indefinitely, thus generating a long 1D-cluster of
entangled photons.

Schwartz and coworkers demonstrated the first proof-of-concept realization
of this proposal in 2016 \citep{Schwartz2016}. They showed that the
entanglement robustness of the 1D - photonic string is mainly determined
by the ratio between the photon radiative time and the spin-precession
time and to a lesser extent also by the ratio between the later and
the confined spin coherence time \citep{Schwartz2016}.

In Ref \citep{Schwartz2016}, the entangler was the spin of the dark
exciton (DE). The short-range electron-hole exchange interaction removes
the degeneracy of the DE even in the absence of external field, therefore
a coherent superposition of the DE eigenstates naturally precesses.
Due to the limited temporal resolution of the silicon avalanche photodetectors
which Schwartz et al used, the spin was re-excited every 3/4 of its
precession period. In this work, we use instead superconducting single
photon detectors with an order of magnitude better temporal resolution.
Therefore we are able to drive the system every 1/4 of the DE precession
period. This leads to photon generation rate which is three fold faster
than previously demonstrated.

We develop a novel experimental and theoretical method for characterizing
the improved cluster state and the spin - multi-photon quantum states
that we generate. Our tomographic method differs from the traditional
method \citep{James2001} in the sense that it enables to measure
the spin that remains in the QD after projecting all the emitted photons.
The method uses time resolved spin-multi-photon correlations for measuring
the quantum state, and for characterizing the periodically used process
map which generates the photonic cluster.

We use a novel gradient descent method to find the process map which
best fits the data in the sense of having maximum likelihood. Our
gradient descent method differs from the standard one in the fact
that we define the gradient relative to a specific non-euclidean metric
which is adapted to the geometry of the set of physical (completely
positive) process maps. This approach is very different from known
algorithms such as projected gradient descent \citep{Goncalves2015,Bolduc2017}.

In the following, we demonstrate our tomographic technique by characterizing
the enhanced gigahertz rate generated cluster state. We show that
as a result of the time reduction between the sequential excitations,
the effect of the DE spin decoherence \citep{Cogan2018} is reduced,
and the robustness of the entanglement in the cluster state increases,
persisting for 6 consecutive photons.

The tomographic method and our experimental results are described
below.

\begin{figure*}
\begin{centering}
\includegraphics[width=1\textwidth]{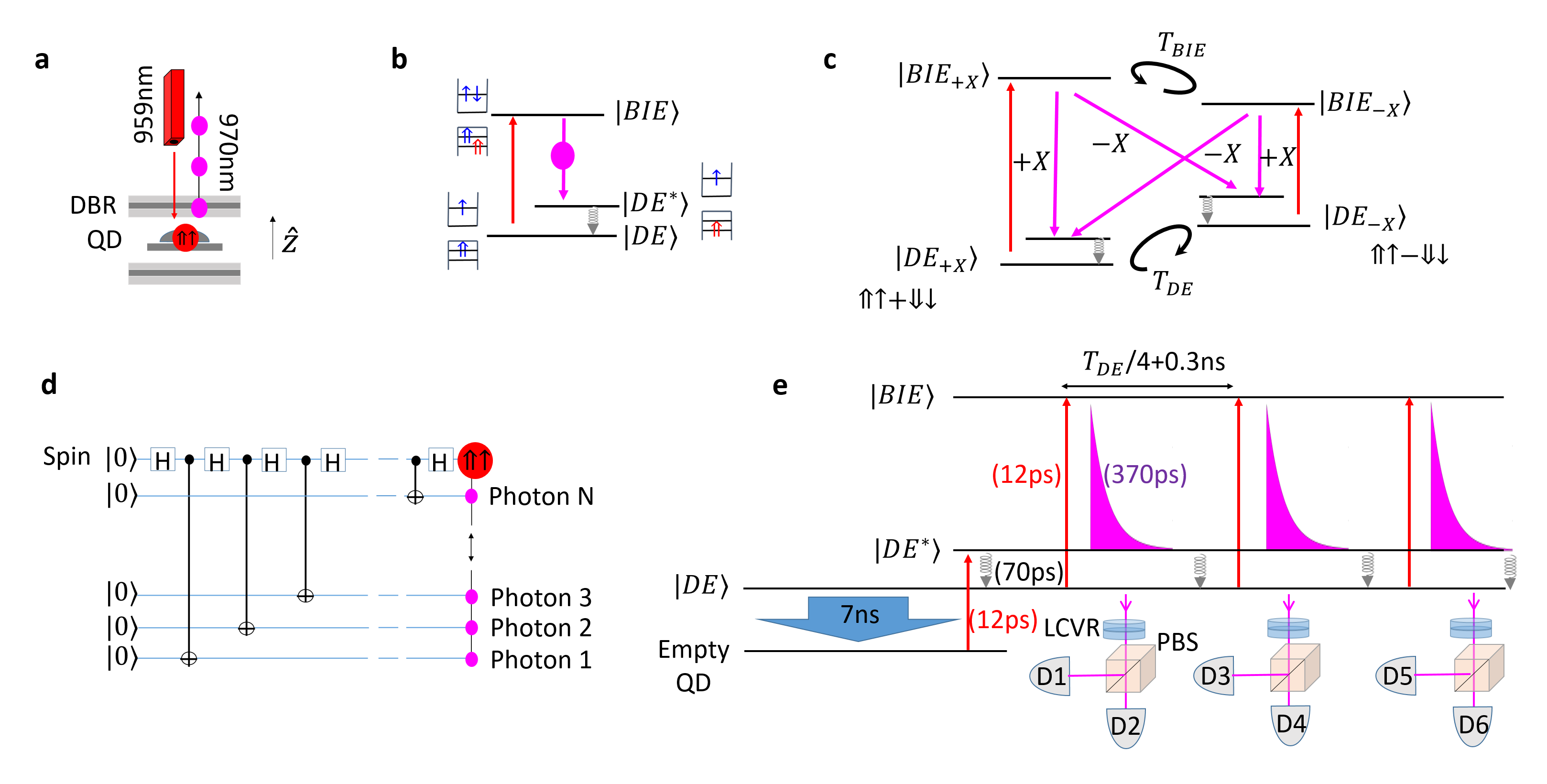}
\par\end{centering}
\caption{\label{fig:Method} a) Schematic description of the QD-based device.
DBR- distributed Bragg reflector. b) The energy levels of the DE -
DE{*} - BIE system and the transitions between these levels. Each
laser pulse (red upward arrow) excites the DE and photogenerates a
BIE. The BIE then decays to an excited DE{*} by emitting a single-photon
(downward pink arrow). The DE{*} then decays to its ground level by
emitting acoustical phonon (curly downward arrow). c) The BIE-DE{*}-DE
energy levels and the polarization-selection-rules for the optical
transitions. Here +X (-X) describes horizontally (vertically) rectilinearly
polarized optical transition. The short-range electron-hole exchange
interaction removes the degeneracy between both the DE and the BIE
two eigenstates. A coherent superposition of the DE (BIE) eigenstates
precesses with a period marked by $T_{DE}$ ($T_{BIE}$). d) The cluster
state repeating protocol contaning Hadamard gates acting on the QD
confined spin followed by a CNOT- two qubit gates, which entangles
the spin polarization with the polarization of an emitted photons.
e) The experimental setup, pulse sequence, photon emissions, and detection
for realizing and characterizing the cluster state. Blue downward
wide-arrow is an optical-depletion 7ns long pulse \citep{Schmidgall2015},
which depletes the QD from carriers. Upward arrows are 12 ps long
$\pi$-area laser pulses, while the downward pink exponential decays
represent the emitted photons. We project the photons' polarization
using liquid-crystal-variable-retarders (LCVRs) and polarizing-beam-splitters
(PBS) while looking for three consecutive photon detections.}
\end{figure*}

\section{Cluster-state Generation - Method}

In the heart of our device is a semiconductor QD. The QD contains
a confined electronic-spin, which serves as the entangler qubit \citep{Loss1998,DiVincenzo_2000,Lindner2009}.
We define the sample growth direction, which is also the QD\textquoteright s
shortest dimension (about 3nm) as the quantization z-axis. The QD
is embedded in a planar microcavity, formed by 2 Bragg-reflecting
mirrors (Fig.~\ref{fig:Method}a), facilitating efficient light-harvesting
by an objective placed above the QD.

The DE is an electron-hole pair with parallel spins \citep{Poem2010,Schwartz2015,Schwartz2015a},
having two total angular momentum states of $\pm2$ as projected on
the QD z-axis. Since the DE optical activity is weak \citep{zielinsky2014},
it has long life and coherence time \citep{Schwartz2015,Cogan2018}.
Upon optical excitation, the DE ($\ket{\Uparrow\uparrow}$) is excited
to form a biexciton (BIE) ($\ket{\Uparrow\Uparrow\uparrow\downarrow}$).
The BIE is formed by pair of electrons in their first conduction-subband
level and 2 heavy holes with parallel spins one in the first and one
in the second valence-subband levels. Fig.~\ref{fig:Method}b and
Fig.~\ref{fig:Method}c schematically describe the DE-BIE energy
level structure, and the selection rules for optical transitions between
these levels, respectively \citep{Bayer2002,Ivchenko2005,Poem2010}.

Each laser pulse (in red) excites the QD confined DE to its corresponding
BIE state. The BIE decays to the DE{*} level within about 370ps by
radiative recombination in which a single photon is emitted (marked
in pink). The DE{*} then decays to the ground DE state by about 70ps
spin-preserving acoustical-phonon relaxation. The energy difference
between the emitted photon and the exciting laser allows us to spectrally
filter the emitted single photons.

The eigenstates of the DE and BIE are given by: $\ket{\pm X_{DE}}=\left(\ket{+Z_{DE}}\pm\ket{-Z_{DE}}\right)/\sqrt{2}$
and $\ket{\pm X_{BIE}}=\left(\ket{+Z_{BIE}}\pm\ket{-Z_{BIE}}\right)/\sqrt{2}$.
The energy differences between these eigenstates are about $1\mu eV$,
smaller than the radiative width of the BIE optical transition ($\simeq3\mu eV$)
and much smaller than the spectral width of our laser pulse ($\simeq100\mu eV$).
It follows that a coherent superposition of the DE (BIE) eigenstates
precesses with a period $T_{DE}=3.1ns$ ($T_{BIE}=5.6ns$), an order
of magnitude longer than the BIE radiative time.

The DE and BIE act as spin qubits. Angular momentum conservation during
the optical transitions between these two qubits imply the following
$\pi$-system selection rules \citep{Cogan2018}:
\begin{align}
\ket{\Uparrow\uparrow} & \stackrel{\ket{+Z}}{\longleftrightarrow}\ket{\Uparrow\Uparrow\downarrow\uparrow},\nonumber \\
\ket{\Downarrow\downarrow} & \stackrel{\ket{-Z}}{\longleftrightarrow}\ket{\Downarrow\Downarrow\uparrow\downarrow}.\label{eq:selection-rules-1-1}
\end{align}

where $\ket{+Z}$ ($\ket{-Z}$) is a right (left) -hand circularly
polarized photon propagating along the +Z-direction. It thereby follows
that a laser pulse polarized $\ket{+X}=\left(\ket{+Z}+\ket{-Z}\right)/\sqrt{2}$
coherently excites a superposition $\alpha\ket{\Uparrow\uparrow}+\beta\ket{\Downarrow\downarrow}$
of the DE spin qubit states to a similar superposition $\alpha\ket{\Uparrow\Uparrow\uparrow\downarrow}+\beta\ket{\Downarrow\Downarrow\uparrow\downarrow}$
of the BIE states. The BIE then radiatively decays into an entangled
spin-photon state $\alpha\ket{\Uparrow\uparrow}\ket{+Z}+\beta\ket{\Downarrow\downarrow}\ket{-Z}$.
Therefore the excitation and photon emission act as a 2-qubit entangling
(CNOT) gate between the spin and the photon.

The method for generating the cluster state is described in Fig.~\ref{fig:Method}d.
The confined DE is resonantly excited repeatedly by a laser pulse
to its corresponding BIE. The BIE decays radiatively by emitting a
photon. The excitation and photon emission act as a two-qubit CNOT
gate which entangles the emitted photon polarization qubit and the
spin qubit, thus adding a photon to the growing photonic cluster.
The excitation pulses are timed such that between the pulses the DE-spin
precesses quarter of its precession period. This temporal precession
can be ideally described as a unitary Hadamard gate acting on the
spin qubit only. The combination of the CNOT 2-qubit gate and the
Hadamard 1-qubit gate forms the basic cycle of the protocol which
when repeated periodically generates the entangled spin + photons
cluster state \citep{Lindner2009}.

For the experimental realization of the cluster protocol and the characterization
of the generated state, we use the experimental setup described in
Fig.~\ref{fig:Method}e. The QD is first optically-emptied from carriers,
making it ready for initialization. The first 7-ns-long optical pulse
(Blue downward arrow) depletes the QD from charges and the remaining
DE \citep{Schmidgall2015}. We then write the DE spin state using
horizontally polarized 12-ps optical $\pi$-pulse to the DE{*} state.
This is possible due to small mixing between the bright exciton (BE)
and the DE \citep{Schwartz2015a}. The pulsed polarization defines
the DE{*} initial spin state. The DE{*} then relaxes to its ground
DE state, making the QD ready for implementing the cluster protocol.
A sequence of resonantly tuned linearly polarized $\pi$-area laser
pulses is then applied to the QD. Each pulse results in the emission
of a photon from the QD's BIE-DE optical transition. During the last
emission, the BIE spin evolution can be conveniently used as a resource
for the DE spin tomography \citep{Cogan2020}.

To characterize the generated multi-qubit quantum state, we project
the polarization of the detected photons on 6 different polarization
states using liquid-crystal-variable-retarders (LCVRs) and polarizing-beam-splitters
(PBSs). We then use highly efficient transmission gratings to spectrally
filter the emitted photons from the laser light. The photons are eventually
detected by 6 efficient (>80\%) fast single-photon superconducting
detectors with temporal resolution of about 30ps.

\section{\label{sec:Measurements}Cluster-state characterization}

The cluster state entanglement robustness is characterized using three
cycles of the repeated protocol. The characterization is done by correlating
one, two, and three detected photon events. In all cases the last
detected photon is used for the tomography of the DE-spin. We use
two different orthogonal, +X and +Y linearly polarized excitation
pulses respectively \citep{Cogan2020}. The +X-polarized laser pulse
promotes the DE state to a similar superposition of BIE states, while
+Y excitation introduces a $\pi/2$ phase shift to the superposition
\citep{Cogan2020}. In addition, we utilize the BIE state evolution
during its radiative decay back to the DE{*} to measure the degree-of-circular-polarization
($D_{cp}$) of the emission as a function of time: 
\begin{equation}
D_{cp}(t)=\frac{P_{+Z}(t)-P_{-Z}(t)}{P_{+Z}(t)+P_{-Z}(t)},
\end{equation}
where $P_{j}$ represents the detected photon polarization-projection
on the j basis. By fitting the measured $D_{cp}(t)$ to a central-spin-evolution-model
that we recently developed for QD confined charge carriers \citep{Cogan2020},
we accurately extract the DE spin state in the time of its excitation.

As we implement the protocol, we perform full tomographic measurements
of the growing quantum state. First, we measure the initialized DE
state. Then we apply one cycle of the protocol and measure the resulting
spin+1photon state. Finally, we apply a second cycle of the protocol
and measure the spin+2photons state.

\begin{figure}
\begin{centering}
\includegraphics[width=1\columnwidth]{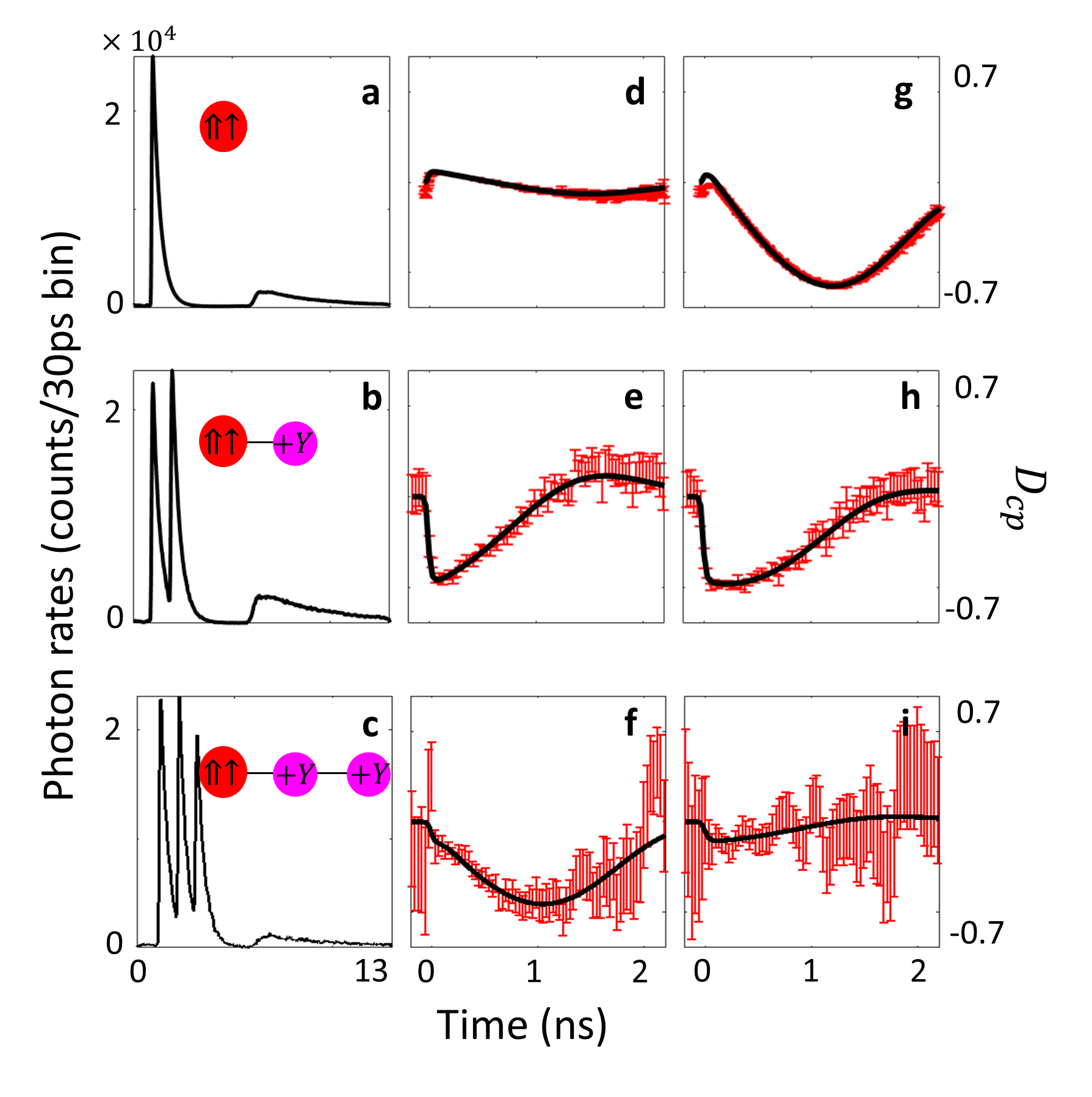}
\par\end{centering}
\caption{\label{fig: Correlations} Example of one-, two-, and three-photon
correlation measurements performed in characterizing the photonic
cluster state. a), b), and c) are time resolved PL measurements of
the BIE-DE optical transition following one-, two- ,and three-pulse
excitations, respectively. d), e), and f) are time resolved degree
of circular polarization measurements $D_{cp}(t)=\frac{P_{+Z}(t)-P_{-Z}(t)}{P_{+Z}(t)+P_{-Z}(t)}$
of the last detected photon, using horizontally polarized (+X) excitation
pulse. g), h), and i) are the same as g), e), and f), using diagonally
linearly polarized excitation pulse (+Y). The red data points describe
the measured data and the solid black line describe the fitted model
calculations (see Appendix). From the fitted model we extract the
spin, spin+1 photon, and spin+2 photon density matrices elements.}
\end{figure}

In Fig \ref{fig: Correlations} we present a small set of the measurements
used to deduce the spin, spin+1photon, spin+2photon quantum states
and the process map of the periodic cycle of the cluster protocol.
The first row shows $D_{cp}$ measurements characterizing the initialized
spin, the second row shows the $D_{cp}$ of the correlated spin+1photon
state after the photon was projected on a +Y polarization basis, and
the third row shows the $D_{cp}$ of the correlated spin+2photons
state after both photons were projected on +Y polarization basis.
Each row differs from the row above it by applying one additional
cycle of the protocol. In all measurements shown in the Fig.~\ref{fig: Correlations},
the DE was initialized to -X state. In each row, the left-panel displays
time resolved PL, the center-panel (right-panel) displays time resolved
$D_{cp}$ measured after +X (+Y) polarized excitation of the final
spin. By fitting the $D_{cp}$ correlation measurements, we extract
the following spin, spin+1photon, and spin+2photon polarization density
matrix elements, respectively: 
\[
[S_{X},S_{Y},S_{Z}]=[-0.73,0.05,0.06]
\]
\[
[P_{Y}^{(1)}S_{X},P_{Y}^{(1)}S_{Y},P_{Y}^{(1)}S_{Z}]=[0.01,0.16,-0.59]
\]

\[
[P_{Y}^{(1)}P_{Y}^{(2)}S_{X},P_{Y}^{(1)}P_{Y}^{(2)}S_{Y},P_{Y}^{(1)}P_{Y}^{(2)}S_{Z}]=[0.03,-0.49,-0.08],
\]
where $P_{j}^{(i)}$represents the polarization projection of the
i'th-photon in the string, on the j polarization base, and $S_{j}$
is the DE-spin polarization, projected on the j base. The typical
measurement uncertainties are about 0.01, 0.02, and 0.04 for the spin,
spin+1photon, and spin+2photon polarization density matrix elements,
respectively.

For a perfect initialization and application of the process we expect
these polarization elements to be {[}-1,0,0{]}, {[}0,0,-1{]}, {[}0,-1,0{]}
respectively. Here, the DE spin is initialized to the $-X$ state
with polarization degree of 0.73, due to the limited efficiency of
the depleting pulse \citep{Schmidgall2015}. After each cycle of the
protocol, the measured $D_{cp}(t)$ is reduced by approximately 20\%,
indicating an exponential decay as expected \citep{Popp_2005,Schwartz2016}.

\begin{figure*}
\begin{centering}
\includegraphics[width=1\textwidth]{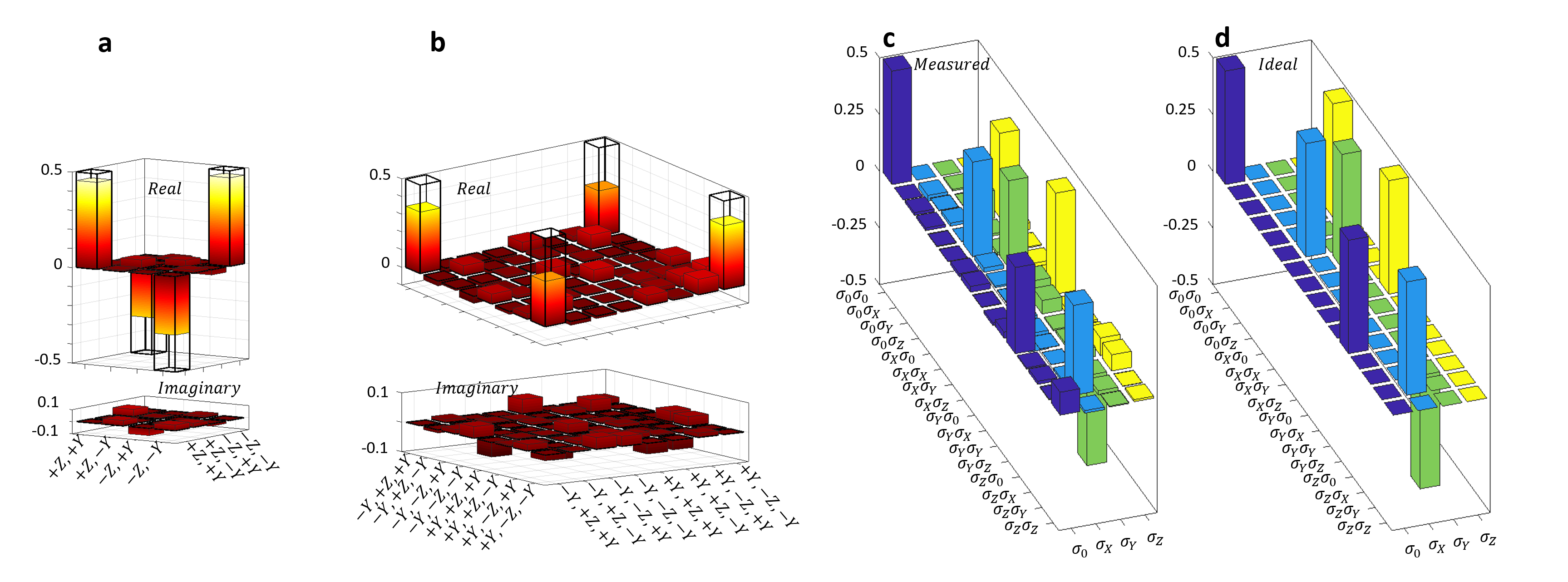}
\par\end{centering}
\caption{\label{fig:Process} a) The measured DE+1photon density matrix. b)
The measured DE+2photons density matrix. In a) and b), the colored
bars represent measured density matrix elements, while the empty bars
represent the ideal maximally entangled 2- and 3-qubit states. The
fidelity of the measured density matrix to the ideal one is $F=0.77\pm0.04$
in a) and $F=0.68\pm0.07$ in b). In both cases, the DE spin is initialized
to the $\protect\ket{-X_{DE}}$ state with measured fidelity of $F=0.86\pm0.02$
and then applying once or twice the process. The process map $\Phi$,
which describes the system's evolution in each cycle of the protocol
is given by $\Phi(\hat{\rho}_{DE})=\protect\underset{\alpha,\beta,\gamma}{\sum}\Phi_{\alpha\beta}^{\gamma}\rho_{\gamma}^{DE}\hat{\sigma}_{\alpha}\hat{\sigma}_{\beta}$,
where $\hat{\rho}_{DE}=\protect\underset{\gamma}{\sum}\rho_{\gamma}^{DE}\hat{\sigma}_{\gamma}$
is the density matrix that describes the input DE state and $\Phi(\hat{\rho}_{DE})$
describes the DE+1photon state after applying the process on the input
DE qubit. Here, $\alpha,\,\beta,$ and $\gamma$ run over $O,\,X,\,Y$,
and $Z$, where $\hat{\sigma}_{0}$ is the identity matrix and $\hat{\sigma}_{X}$,
$\hat{\sigma}_{Y}$, and $\hat{\sigma}_{Z}$ are the corresponding
Pauli matrices. The 64 real parameters $\Phi_{\alpha\beta}^{\gamma}$
, thus fully specify $\Phi$. c) The process map $\Phi_{\alpha\beta}^{\gamma}$
as measured by the quantum process tomography. The matrix elements
of $\Phi_{\alpha\beta}^{\gamma}$ are presented such that the rows
correspond to the indices $\alpha\beta$ of the DE+1photon output
state, and the columns correspond to the index $\gamma$ of the input
DE state. d) The process map $\Phi_{\alpha\beta}^{\gamma}$, calculated
assuming the ideal protocol (CNOT and Hadamard gates), as in Fig.~\ref{fig:Method}d. }
\end{figure*}

The measured correlations are used to infer directly the spin+1photon
2qubit state, obtained by applying one cycle of the protocol, the
spin+2photon state obtained by applying two cycles of the protocol,
and a full process tomography of the periodic cycle of the cluster
protocol.

To measure the spin+1photon density matrix (displayed in Fig.~\ref{fig:Process}a),
we use a set of 12 DE-photon $D_{cp}(t)$ correlation measurements.
Two of those measurements are displayed in Fig.~\ref{fig: Correlations}e
and Fig.~\ref{fig: Correlations}h. The measured density matrix has
fidelity of $F=0.77\pm0.04$ with the maximally entangled Bell state.
To measure the spin+2photons 3-qubit density matrix (displayed in
Fig.~\ref{fig:Process}b) we use a set of 72 $D_{cp}(t)$ DE-2photon
correlation measurements. Two of those measurements are displayed
in Fig.~\ref{fig: Correlations}f and \ref{fig: Correlations}i.
The measured spin+2photons density matrix has fidelity of $0.68\pm0.07$
with the maximally entangled 3-qubit state.

We produce the cluster state by repeatedly applying the same process,
as shown in the protocol of Fig.~\ref{fig:Method}d. As a result,
one can fully characterize the cluster state for any number of qubits
if the single-cycle process-map is known \citep{Schwartz2016}. Ideally,
the process-map contains a CNOT and a Hadamard gate. It maps the 2$\times$2
spin qubit density matrix into a 4$\times$4 density matrix representing
the entangled spin-photon state. The process map $\Phi$ can be fully
described by a 4$\times$16 positive and trace-preserving map with
64 real matrix elements.

We use the convention $\Phi(\hat{\rho}_{DE})=\underset{\alpha,\beta,\gamma}{\sum}\Phi_{\alpha\beta}^{\gamma}\rho_{\gamma}^{DE}\hat{\sigma}_{\alpha}\hat{\sigma}_{\beta}$
, where $\hat{\rho}_{DE}=\underset{\gamma}{\sum}\rho_{\gamma}^{DE}\hat{\sigma}_{\gamma}$
is the density matrix that describes the input DE state and $\Phi(\hat{\rho}_{DE})$
describes the DE+1photon state after one application of the process
to the input DE state. The sums are taken over $\alpha,\beta,\gamma=O,X,Y,Z$,
where $\hat{\sigma}_{0}$ is the identity matrix and $\hat{\sigma}_{X}$,$\hat{\sigma}_{Y}$,$\hat{\sigma}_{Z}$
are the corresponding Pauli matrices. The 64 real parameters $\Phi_{\alpha\beta}^{\gamma}$
thus fully specify $\Phi$.

Fig.~\ref{fig:Process}c shows the results of the full-tomographic
measurements of the process map. For acquiring these measurements,
we initialize the DE-spin-state to six different states from three
orthogonal bases \citep{Schwartz2015a}. For each of those 6 states
we use 2 $D_{cp}(t)$ single-photon measurements like the measurements
displayed in Fig.~\ref{fig: Correlations}d and Fig.~\ref{fig: Correlations}g
for tomography. Then we apply one cycle of the protocol for each initialization
and measure the resulting spin+1photon states by projecting the first
photon on different orthogonal polarization bases \citep{James2001}
and correlating it with the $D_{cp}(t)$ of the second photon. For
characterizing each of those 6 spin+1photon states, we use 12 $D_{cp}(t)$
2-photon correlations measurements like the ones presented in Fig.~\ref{fig: Correlations}e
and Fig.~\ref{fig: Correlations}h.

To obtain the physical process map that best fits our measured results,
we use a specifically developed edge-sensitive twisted-gradient-descent
minimization method (See Appendix). It is well known that the space
of physical completely-positive (CP) maps can be identified with a
cone-like-space where any unitary process sits on an extremal ray
of the cone. To find the best CP fit of $\Phi$ therefore requires
minimizing a known function $F$ (representing minus log likelihood)
over this cone. Gradient descent tries to find the minimum of a function
$F(x)$ by going roughly along gradient lines of $F$. Our newly developed
approach uses an edge-sensitive twisted-gradient-descent in order
to prevent our gradient descent solution from getting stuck in the
boundary of the physically allowed cone-like region.

Fig.~\ref{fig:Process}c-d presents the physical CP-process-map obtained
using this method. We compare the acquired physical process with the
ideal unitary process of the cluster protocol. The fidelity \citep{Jozsa1994,Schwartz2016}
between the two processes is 0.83. The obtained fidelity is higher
than in the previous demonstration \citep{Schwartz2016}. The higher
fidelity is attributed to the 3-fold shorter time between the excitations,
which reduces the influence of the DE decoherence during its precession.
The relatively high fidelity to the ideal protocol indicates that
our device can deterministically generate photonic cluster states
of high quality, thereby providing a better resource for quantum information
processing.\\

\section{Discussion}

\begin{figure}
\begin{centering}
\includegraphics[width=1\columnwidth]{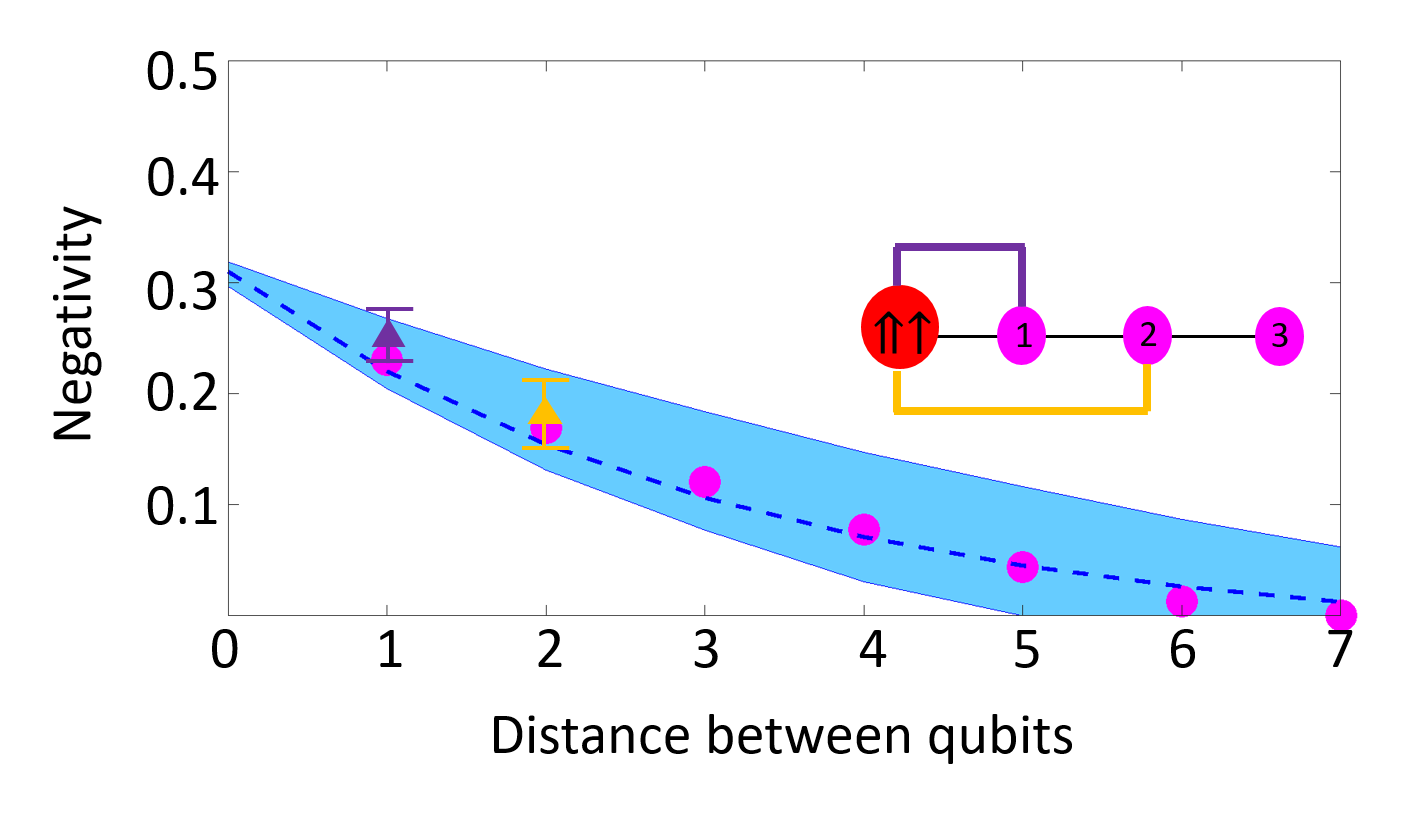}
\par\end{centering}
\caption{\label{fig:EL}The localizable entanglement (LE - pink circles) in
the generated state vs. the distance between two qubits in the string.
The pink circles correspond to the LE between qubits m and m + d in
the state of DE+Nphotons obtained using the measured process map.
The dashed lines represent the best fit to an exponential decay law
$N(d)=N_{nn}\exp\left(-(d-1)/\zeta_{LE}\right)$, where $N_{nn}$
is the negativity \citep{Peres1996} between nearest-neighbor qubits,
d is the distance between the qubits, and $\zeta_{LE}$ is the characteristic
decay-length of the LE. The light-blue shaded area represents one
standard deviation of the uncertainty in the measurements determining
the process map. The purple and yellow data points represent the directly
measured LE of the DE and the first emitted photon in a two- and a
three-qubit string, respectively, acquired from the complete tomographic
measurements of those states presented in Fig.~\ref{fig:Process}c
and d.}
\end{figure}

We characterize the robustness of the entanglement in the 1D cluster-state
using the notion of localizable entanglement (LE) \citep{Verstraete2004}.
The LE is the negativity \citep{Peres1996} between two qubits in
the cluster after all the other qubits are projected onto a suitable
polarization-basis. The LE decays exponentially with the distance
between the qubits \citep{Popp_2005,Schwartz2016}: 
\begin{equation}
N(d)=N_{nn}\exp\left(-(d-1)/\zeta_{LE}\right),\label{eq:LE}
\end{equation}
where $N_{nn}$ is the negativity between nearest-neighbor qubits,
d is the distance between the qubits, and $\zeta_{LE}$ is the characteristic
decay-length of the LE. In Fig.~\ref{fig:EL}, we plot using pink
circles the LE in the state of a spin+Nphotons, obtained from the
measured process map, as a function of the distance between two qubits
in the string. As expected, the LE in the 1D cluster state decays
exponentially with the distance between the two qubits \citep{Popp_2005}.
Fig.~\ref{fig:EL} shows that the entanglement in the cluster persists
up to up to six photons. This presents an improvement over Ref.~\citep{Schwartz2016},
resulting from the reduction in the DE spin decoherence between the
optical pulses.

The negativity between nearest neighbor and next nearest neighbor
2 qubits can be directly obtained from our quantum state tomography.
The entanglement between the DE and the photon emitted after one cycle
of the protocol is obtained from the density matrix of the DE+1photon
in Fig.~\ref{fig:Process}a which has negativity of $N=0.27\pm0.03$.
This negativity is marked as a purple data point in Fig.~\ref{fig:EL}.
Similarly, the spin+2photons 3-qubit state resulted from application
of two cycles of the protocol is represented by the 3-qubit density
matrix, displayed in Fig.~\ref{fig:Process}b. The negativity of
the density matrix of the two external qubits, after projecting the
central qubit on the X polarization basis is $N=0.18\pm0.05$, marked
by the yellow  point in Fig.~\ref{fig:EL}. \\

In summary, we demonstrate a gigahertz-rate deterministic generation
of entangled photons in a cluster state, which is 3-times faster than
previously demonstrated. We developed a novel method for spin - multi
photon quantum state tomography and for characterizing the periodic
process map which generates the photonic cluster state. Using this
method we show that the enhanced cluster generation rate also improves
the robustness of the entanglement in the generated multi -photon
state. The measured process map has fidelity of 0.83 to the ideal
one, and the entanglement in the cluster state persists up to 6 consecutive
qubits. Our studies combined with further feasible optimizations of
the device may lead to implementations of quantum communication and
efficient distribution of quantum entanglement between remote nodes.

\section*{acknowledgments}

The support of the Israeli Science Foundation (ISF), and that of the
European Research Council (ERC) under the European Union\textquoteright s
Horizon 2020 research and innovation programme (Grant Agreement No.
695188) are gratefully acknowledged.

\bibliographystyle{aipnum4-1}

\appendix
\maketitle

\section{the process likelihood function}

\label{df0}

The basic (single cycle) process taking the QD-spin into a spin plus
emitted photon is described by a process map $\Phi$ taking one-qubit
state into a two-qubit state. Expanding these quantum states in terms
of Pauli matrices allows writing the map as 
\[
\rho=\sum r_{\mu}\sigma^{\mu}\mapsto\Phi(\rho)=\sum r_{\mu}\phi_{\nu\lambda}^{\mu}\sigma^{\nu}\otimes\sigma^{\lambda}.
\]
Here $\phi_{\nu\lambda}^{\mu}$ with $\mu\nu\lambda\in\{0,x,y,z\}$
are 64 real coefficient which define the process map $\Phi$.

The trace preserving condition $Tr\Phi(\rho)=Tr\rho\forall\rho$ fixes
4 out of the 64 coefficient $\phi_{\nu\lambda}^{\mu}$ namely it requires
that $\phi_{00}^{\mu}=\frac{1}{2}\delta_{\mu0}$. We wish to estimate
the other 60 parameter by finding the best fit to the experimental
data.

In an experiment where the initial spin was $\vec{s}$ and the emitted
photon was projected on state of spin \footnote{we identify H,B,R polarizations with $\hat{x},\hat{y},\hat{z}$ respectively.}
$\vec{p}$ we would ideally expect the measured rate $R$ of photon
emission and final spin $\vec{S}$ to satisfy $\phi_{\nu\lambda\mu}s_{\mu}p_{\lambda}=R\;S_{\nu}$.
Here $s,p,S$ are four-vectors whose zeroth component equals 1 and
their spatial components are $\vec{s},\vec{p},\vec{S}$. Repeating
such measurement for six different spin polarizations $\vec{s}$ and
six different photon polarizations $\vec{p}$ gives $6\times6\times4=144$
equations for the $60$ unknown components of $\Phi$.

A straightforward approach to determine $\Phi$ is then to use a least
square method minimizing the expression $F_{0}=\sum_{p,s,\nu}\frac{1}{\Delta_{ps\nu}^{2}}\left[\phi_{\nu\lambda\mu}s_{\mu}p_{\lambda}-R\;S_{\nu}\right]^{2}.$
(Essentially representing minus log likelihood.) Here $\Delta_{ps\nu}$
are the error estimates of the corresponding measurement. 
A slightly more sophisticated approach would recall that appearance
of each of the six initial polarizations $\vec{s}$ in 24 different
sums, can potentially make the errors correlated. One way to take
this into consideration is to use a more complicated square sum constructed
using a non-diagonal covariance matrix. A completely equivalent method
consists of defining new variables $\vec{s'}$ representing the 'true'
initial polarization of the DE and using the new sum of square error
function $\tilde{F}_{0}=\tilde{F}_{0}(\phi,s')$ defined by \footnote{The error estimate $\Delta_{ps\nu}$ appearing in $\tilde{F}_{0}$
is constructed in the standard way from the error estimates $\Delta R$
and $\Delta S$. The error $\Delta s$ is now taken care of by the
second term (and $\Delta p$ is negligible anyway).} 
\[
\tilde{F}_{0}(\phi,s')=\sum_{p,s,\nu}\frac{1}{\Delta_{ps\nu}^{2}}\left[\phi_{\nu\lambda\mu}s'_{\mu}p_{\lambda}-R\;S_{\nu}\right]^{2}+\sum_{s}\sum_{i=1}^{3}\left(\frac{s'_{i}-s_{i}}{\Delta s_{i}}\right)^{2}
\]
Minimizing this expression with respect to $s'$ yields a standard
square sum $F_{0}(\phi)$ corresponding to the correct non-diagonal
covariance matrix. We looked for a minimum of $\tilde{F}_{0}(\phi,s')$
with respect to both the process map $\phi$ and the unknown initialization
polarizations $s'$.

\section{Completely positive condition and twisted gradient descent.}

It is well known that to be physically acceptable, a process map $\Phi$
must be completely positive (CP). A process map is CP iff the associated
Choi matrix which may be defined (up to unimportant normalization
factor) by 
\begin{equation}
C_{\Phi}=\sum\phi_{\nu\lambda}^{\mu}\;\overline{\sigma_{\mu}}\otimes\sigma^{\nu}\otimes\sigma^{\lambda}\label{choi}
\end{equation}
is positive (semi-definite) $C_{\Phi}\geq0$. The bar over $\sigma_{\mu}$
denotes complex conjugation and we make no distinction here between
lower and upper indices. In general the Choi matrix $C_{\Phi}$ is
a (complex) hermitian $8\times8$ matrix which may be used as an alternative
description of $\Phi$.

Simple minded naive minimization of $F_{0}$ leads to a process map
which is not CP and hence not physically acceptable. To understand
why this happens, recall that our process is very close to an idealized
process which is unitary. Any unitary process is an extreme point
of the cone of CP-maps and has a rank-1 Choi matrix. In other words,
for a unitary process 7 out of the 8 eigenvalues of the Choi matrix
vanish. Our $\Phi$ being close to unitary has therefore 7 very small
Choi matrix eigenvalues. It is thus not surprising that very small
experimental errors can lead us to estimate some of these eigenvalues
as negative, in contradiction with the CP condition. (Had our $\Phi$
been equal to the ideal unitary map, a small random error in each
eigenvalue would lead to non CP map with probability $\frac{127}{128}=0.992$)
In other words, the encountered difficulty is actually a good sign
indicating that our process has quite low decoherence.

To find the best CP fit of $\Phi$ therefore requires minimizing a
known function $F$ (representing minus log likelihood) over the subset
of CP-maps, which as explained above may be identified (through the
Choi representation) with the cone of positive matrices. This is closely
related to the extensively studied field of convex optimization. We
have not found however in the convex optimization literature a method
which looks to exactly fit our problem. We have therefore devised
a method of our own (explained below) which is a variant of the well
known gradient descent.

Gradient descent tries to find the minimum of a function $F(x)$ by
going roughly along gradient lines of $F$. It corresponds to (numerically
discretized) solution to the equation $\frac{d}{dt}x^{i}=-g^{ij}\partial_{j}F(x)$.
The (inverse) metric tensor $g^{ij}$ is often taken to be the standard
euclidean metric $\delta_{ij}$. Such choice is not mandatory and
in fact one may choose any (positive) metric. We suggest to use a
smarter choice of the metric in order to prevent our gradient descent
solution from getting stuck in the boundary of the physically allowed
region.

It is easiest to understand our approach by considering minimization
over a simple region like $\{(x,y)\in\mathbb{R}^{2}\mid x,y\geq0\}$.
In this case our approach to minimizing $F(x,y)$ would correspond
to using iteration steps with $(\Delta x,\Delta y)\propto(x\partial_{x}F,y\partial_{y}F)$.
Assuming that both derivatives of $F$ are $O(1)$, one sees that
if our approximate estimate $(x,y)$ of the minimizer is very close
to one of the boundaries e.g. if it has $x\ll1$ and $y=O(1)$ then
the next step $(\Delta x,\Delta y)$ would adapt to this fact by being
almost parallel to the $y$-axis. One can then go a long $\sim O(1)$
distance along this direction without crossing the boundary. This
is in contrast to hitting the boundary after a distance $O(x)\ll1$
which would result from using the standard metric $g_{ij}=\delta_{ij}$.

Since the CP - condition is easier to formulate in terms of Choi matrices,
let us consider the (square sum) function we want to minimize as a
function $F(A)$ over the (real vector space) of hermitian matrices
\footnote{It is convenient to extend $F$ to arbitrary matrices by defining
$F(A)=F(\frac{1}{2}(A+A^{\dag}))$.}. The standard euclidean gradient of $F$ may be identified with the
matrix $\nabla F$ whose elements are $(\nabla F)_{ij}=\frac{\partial F}{\partial a_{ji}}$.
(This relation may be a bit confusing since the elements $a_{ij}$
of $A$ are complex.) An equivalent and possibly more rigorous definition
starts by expressing $A$ as $A=\sum a_{\alpha}\Xi_{\alpha}$ where
$a_{\alpha}\in\mathbb{R}$ and $\{\Xi_{\alpha}\}$ are some basis
for the space of Hermitian matrices which is orthonormal in the sense
$Tr(\Xi_{\alpha}\Xi_{\beta})=N\delta_{\alpha\beta}$ with some normalization
$N$. Note that our $A$ being a process Choi matrix, is already given
to us in such a form by Eq.(\ref{choi}). One can then write $\nabla F=\sum\Xi_{\alpha}\partial_{\alpha}F$.

The gradient we use corresponds to a non-flat riemanian metric and
may be expressed as $\tilde{\nabla}F=\sqrt{A}(\nabla F)\sqrt{A}$
(where $\nabla F$ is as above). We therefore look for a minimum of
$F$ over the set of positive (semi-definite) matrices by using a
gradient descent step of the form 
\[
A\mapsto A+\Delta A,\;\;\;\;\;\Delta A=-q\sqrt{A}(\nabla F)\sqrt{A}
\]
Here $q>0$ is a scalar chosen so that $F(A+\Delta A)$ is minimal
under the constraint $A+\Delta A\geq0$ . Note that if $\nabla F=O(1)$
then the positivity constraint allows $q$ to remain $O(1)$ even
if $A$ is very close to the boundary of the cone of positive matrices.
All our process map estimations were obtained using this minimization
scheme which we implemented using Mathematica$^{TM}$.

Although the basic method described above works reasonably well, we
found that some extra improvement \footnote{Euclidean gradient descent fails even if one includes similar improvement
in it.} is gained if after each step we push $A$ slightly away from the
boundary of the allowed region by updating it as $A\rightarrow(1-\varepsilon)A+\frac{1}{2}\varepsilon I$
with $\varepsilon\ll1$. (We suspect that the need for this step might
be related to the finite precision of the numerical calculations.)
We increase or decrease $\varepsilon$ dynamically during the computation,
depending on the performance of previous iteration step. 
In practice $\varepsilon$ ranged between $10^{-4}$ and $10^{-10}$
and scaled roughly as 2-3 times the minimal eigenvalue of $A$.

The modified gradient descent method described here is quite general
and can be applied to any $F(A)$. In practice, our $F(A)$ was of
the form $F(A)=F_{0}(A)+\sum_{\mu=0}^{3}\lambda_{\mu}Tr(A(\sigma_{\mu}\otimes I\otimes I))$
where $F_{0}(A)$ is the square sum described in subsection \ref{df0},
and the second term consists of 4 lagrange multipliers required to
enforce the normalization condition \footnote{The need for lagrange multiplier is a cost we pay for using non flat
metric. In flat metric, the constraints may be solved trivially.} $\phi_{00}^{\mu}=\frac{1}{2}\delta_{\mu0}$. The values of the multipliers
$\lambda_{\mu}$ at each iteration step are easily determined numerically
by requiring that $\Delta A=-q\sqrt{A}(\nabla F)\sqrt{A}$ does not
break the normalization condition. This amount to demanding the partial
trace $Tr_{2,3}(\sqrt{A}(\nabla F)\sqrt{A})$ to vanish, which is
just a linear set of equations for $\lambda_{\mu}$.
\end{document}